\documentclass[twocolumn,showpacs,prb,aps,amssymb]{revtex4}
\usepackage{graphicx}
\catcode`\@=11
\def\simle{\mathrel{\mathpalette\@versim<}}   
\def\simge{\mathrel{\mathpalette\@versim>}}   
\def\@versim#1#2{\lower2.5pt\vbox{\baselineskip0pt \lineskip-.5pt
   \ialign{$\m@th#1\hfil##\hfil$\crcr#2\crcr\sim\crcr}}}
\newcommand{\mib}[1]{\mbox{\boldmath $#1$}} 

\begin{document}

\title{
Orbital and spin interplay in spin-gap formation 
in pyroxene $A$TiSi$_2$O$_6$ ($A$=Na, Li)
}
\author{
Toshiya Hikihara and Yukitoshi Motome}
\affiliation{
RIKEN, 2-1 Hirosawa, Wako, Saitama 351-0198, Japan 
}
\date{\today}

\begin{abstract}
Interplay between orbital and spin degrees of freedom is 
theoretically studied for the phase transition 
to the spin-singlet state with lattice dimerization 
in pyroxene titanium oxides $A$TiSi$_2$O$_6$ ($A$=Na, Li). 
For the quasi one-dimensional spin-$1/2$ systems, 
we derive an effective spin-orbital-lattice coupled model 
in the strong correlation limit with explicitly taking account of 
the $t_{2g}$ orbital degeneracy, and investigate the model 
by numerical simulation as well as the mean-field analysis. 
We find a nontrivial feedback effect between orbital 
and spin degrees of freedom; 
as temperature decreases, development of antiferromagnetic spin correlations 
changes the sign of orbital correlations from antiferro to ferro type, 
and finally the ferro-type orbital correlations induce the dimerization 
and the spin-singlet formation. 
As a result of this interplay, 
the system undergoes a finite-temperature transition 
to the spin-dimer and orbital-ferro ordered phase 
concomitant with the Jahn-Teller lattice distortion. 
The numerical results for the magnetic susceptibility show 
a deviation from the Curie-Weiss behavior, and 
well reproduce the experimental data. 
The results reveal that the Jahn-Teller energy scale is considerably small 
and the orbital and spin exchange interactions play a decisive role 
in the pyroxene titanium oxides.
\end{abstract}


\maketitle

\section{Introduction}
\label{sec:introduction}

Orbital degree of freedom has attracted much attention 
since it plays key roles in electronic properties 
of transition metal compounds.\cite{Tokura2000,Imada1998} 
The orbital degree of freedom couples with the Jahn-Teller (JT) lattice 
distortion, and  in many compounds, the energy scale of 
the JT stabilization energy is larger than that of 
the spin and orbital exchange interactions.\cite{Kugel1982} 
A typical example is the mother compound of 
colossal magnetoresistive (CMR) manganites, LaMnO$_3$: 
The JT energy scale is $\sim 0.1$-$1$eV and the spin exchange energy scale is 
$\sim 10$meV. 
Consequently, the orbital-lattice transition occurs at a much higher 
temperature ($\sim 800$K) than the antiferromagnetic (AF) transition 
temperature ($\sim 140$K).\cite{Wollan1955} 
In these systems, the orbital-lattice physics is dominant, and 
the orbital and lattice orderings modify effective spin exchange interactions 
to lead a magnetic ordering in a secondary effect. 
Moreover, the JT distortion suppresses 
fluctuation effects in orbital and spin degrees of freedom. 
Hence, a large JT coupling masks bare interplay between spin and orbital 
in many real materials. 

Quantum and thermal fluctuations 
in the competition between the spin and orbital exchange interactions 
may yield novel phenomena, 
and therefore, it is highly desired to explore systems 
in which the genuine spin-orbital interplay appears explicitly.
One of promising candidates is 
the $t_{2g}$ electron system such as titanium and vanadium oxides. 
In the $t_{2g}$ systems, the JT coupling becomes smaller than in the $e_g$ 
systems such as CMR manganites because spatial shapes of $t_{2g}$ orbitals 
avoid directions of surrounding ligands in the octahedral positions. 
For instance, in vanadium perovskite oxides $A$VO$_3$ ($A$=La, Ce), 
the orbital-lattice transition temperature becomes even lower than 
the AF one,\cite{Miyasaka2003} and novel interplay between spin 
and orbital is proposed theoretically.\cite{Motome2003,Khaliullin2001} 

\begin{figure}
\caption{
(a) Lattice structure of pyroxene oxides $A$TiSi$_2$O$_6$ ($A$=Na, Li).
Chains of TiO$_6$ octahedra are separated by SiO$_4$ tetrahedra.
(b) Skew edge-sharing chain structure of TiO$_6$ octahedra.
Balls in the center of each octahedron denote Ti cations, and 
oxygen ions are on the corners of the octahedra.
The octahedra share their edges in the $xy$ and $yz$ planes alternatively. 
White objects with four lobes denote 
$d_{xy}$ and $d_{yz}$ orbitals of $t_{2g}$ electrons, and 
$t^{11}$ ($t^{22}$) is the $\sigma$-bond transfer integrals 
between $d_{xy}$ ($d_{yz}$) orbitals. 
See the text for details. 
} 
\label{fig:structure}
\end{figure}

Pyroxene titanium oxides $A$TiSi$_2$O$_6$ ($A$=Na, Li) 
are typical examples of the $t_{2g}$ electron systems 
where such interplay between spin and orbital is expected.\cite{Isobe2002}  
The lattice structure of these compounds consists of 
characteristic one-dimensional (1D) chains of 
skew edge-sharing TiO$_6$ octahedra as shown in Fig.~\ref{fig:structure} (a). 
The TiO$_6$ chains are bridged and well separated by SiO$_4$ tetrahedra, 
and therefore, interchain couplings are considered to be much weaker than 
intrachain interactions. 
In each TiO$_6$ chain, as shown in Fig.~\ref{fig:structure} (b), 
the edges of octahedra in the $xy$ and $yz$ planes 
are alternatively shared between neighboring octahedra, 
which leads to the zig-zag structure. 
Each magnetic Ti$^{3+}$ cation has one $d$ electron 
in these insulating materials. 
Hence, we may consider that a quasi 1D spin-$1/2$ system is realized. 

Pyroxene titanium oxides show a peculiar phase transition. 
The magnetic susceptibility shows a sharp drop at $T_{\rm c} = 210$K 
and $230$K in NaTiSi$_2$O$_6$ and LiTiSi$_2$O$_6$, respectively, 
which indicates the transition to a spin-singlet state 
with a finite spin gap.\cite{Isobe2002} 
Below $T_{\rm c}$, a dimerization of 
Ti-Ti distances along the chain was observed 
by the X-ray scattering.\cite{Ninomiya2003} 
These remind us of a spin-Peierls transition.\cite{Boucher1996} 
However, above $T_{\rm c}$, the magnetic susceptibility 
shows an unusual temperature dependence 
which clearly deviates from that of other spin-Peierls compounds. 
In spin-Peierls systems, the dimerization is caused by 
the magnetoelastic coupling, and therefore, 
the transition occurs when short-range spin correlations develop enough 
to drive the lattice dimerization. 
This development of spin correlations is manifested in a broad peak of 
the magnetic susceptibility above $T_{\rm c}$, 
and the peak temperature gives 
a rough estimate of the spin exchange energy scale. 
Contrary to this conventional behavior, 
the magnetic susceptibility of the pyroxene compounds at high temperatures 
increases as temperature decreases, and suddenly drops at $T_{\rm c}$ 
without a clear formation of the broad peak.\cite{Isobe2002} 
This suggests a breakdown of the simple spin-Peierls picture 
in these pyroxene compounds. 

For the peculiar transition to the spin-singlet state, 
an importance of the $t_{2g}$ orbital degree of freedom 
has been pointed out. 
\cite{Isobe2002} 
Theoretically, a scenario of the orbital-driven spin-singlet formation 
has been explored in spin-orbital coupled models 
for $t_{2g}$ electron systems.  
\cite{Katoh1999} 
It was proposed that the orbital ordering may modify 
effective spin exchange interactions and induce the spin-singlet formation. 
However, models were considered only for systems 
with corner-sharing octahedra, and hence, 
it is unclear that the argument is applicable to 
the present pyroxene systems with the edge-sharing octahedra. 
A similar scenario has been 
proposed for the present systems,\cite{Konstantinovic2004}
however, the analysis was heuristic and not sufficient 
to conclude the mechanism of the finite-temperature transition. 
In order to clarify 
the nature of the transition and the low-temperature phase, 
we need more elaborate analysis.
In particular, it is indispensable to investigate 
thermodynamic properties on the basis of a microscopic model. 

In the present study, we will theoretically explore 
the mechanism of the unusual phase transition 
in the pyroxene titanium oxides $A$TiSi$_2$O$_6$ 
explicitly taking account of the $t_{2g}$ orbital degree of freedom. 
We derive an effective spin-orbital-lattice 
coupled model in the strong correlation limit, 
and investigate thermodynamics 
as well as the ground-state properties of the model 
applying the numerical quantum transfer matrix method and  
mean-field-type arguments. 
As a result, we find that interesting interplay 
between orbital and spin degrees of freedom occurs in the system, 
and the interplay gives a comprehensive understanding of 
the peculiar properties of the pyroxene compounds: 
Although both orbital and spin correlations are 
antiferro type and compete with each other at high temperatures, 
development of AF spin correlations with decreasing temperature 
yields a sign change of orbital correlations from antiferro to ferro type. 
After the sign change, the ferro-type orbital correlations 
grow with the antiferro-type spin correlations,
and finally, induce the spin-singlet formation with a dimerization.
Furthermore, the nontrivial temperature dependence of orbital correlations 
modifies an effective magnetic coupling, and 
results in a non Curie-Weiss behavior of the magnetic susceptibility. 
We show that our model with realistic parameters reproduces
the peculiar temperature dependence of the magnetic susceptibility 
in experiment. 

The organization of this paper is as follows.
In the following section \ref{sec:model}, 
applying the strong-coupling approach, 
we derive an effective spin-orbital-lattice coupled model 
for the $t_{2g}$ pyroxene compounds $A$TiSi$_2$O$_6$. 
In Sec.~\ref{sec:MF}, 
we discuss properties of the system in the ground state 
and in the high-temperature limit using mean-field type arguments. 
In Sec.~\ref{sec:results}, 
we study thermodynamic properties of the effective model 
by numerical simulations. 
We describe the method in Sec.~\ref{sec:method}. 
Section \ref{sec:numerical results} shows the numerical results 
including quantitative comparisons with the experimental data.
Finally, Section \ref{sec:summary} is devoted to 
summary and concluding remarks.

\section{Model Hamiltonian}
\label{sec:model}

In this section, we derive an effective spin-orbital-lattice coupled model 
for the pyroxene oxides $A$TiSi$_2$O$_6$, whose Hamiltonian reads 
\begin{equation}
\mathcal{H} = \mathcal{H}_{\rm so} + \mathcal{H}_{\rm JT} 
+ \mathcal{H}_\perp.
\label{eq:H}
\end{equation}
The first term describes the intersite exchange interactions 
in spin and orbital degrees of freedom, and
the second term includes the Jahn-Teller type orbital-lattice coupling. 
These two are defined within each 1D chain. 
The third term describes the interchain coupling. 

The spin-orbital exchange Hamiltonian $\mathcal{H}_{\rm so}$ 
is derived from a $t_{2g}$ multiorbital Hubbard model 
by using the perturbation in the strong correlation limit.
\cite{Kugel1982,Kugel1975} 
The skew structure shown in Fig.~\ref{fig:structure} (b) 
distorts TiO$_6$ octahedra in the form that four Ti-O bonds 
in which the oxygen ions are shared with 
neighboring TiO$_6$ octahedra are longer than 
the rest two Ti-O bonds in each octahedron.\cite{Ninomiya2003} 
The distortion leads to the splitting of threefold $t_{2g}$ levels 
into a low-lying doublet with $d_{xy}$ and $d_{yz}$ orbitals and 
a single $d_{zx}$ level 
when we take the coordinates 
as shown in Fig.~\ref{fig:structure} (b). 
The splitting is estimated as $\sim 300$meV 
in a pyroxene vanadium oxide which has a similar lattice structure.
\cite{Millet1999} 
Because of this level splitting, it is reasonable to start from 
the 1D Hubbard model with twofold degeneracy of $d_{xy}$ and $d_{yz}$ orbitals 
with tracing out the higher $d_{zx}$ level. 
The Hamiltonian is given in the form 
\begin{eqnarray}
\mathcal{H}_{\rm Hub} &=& \sum_{i,j} \sum_{\alpha,\beta} \sum_{\tau}
( t_{ij}^{\alpha\beta} 
c_{i \alpha \tau}^{\dagger} c_{j \beta \tau} + {\rm H.c.} )
\nonumber
\\
&+& \frac12 \sum_i \sum_{\alpha\beta,\alpha'\beta'} \sum_{\tau\tau'} 
U_{\alpha\beta,\alpha'\beta'}
c_{i\alpha\tau}^{\dagger} c_{i\beta\tau'}^{\dagger}
c_{i\beta'\tau'} c_{i\alpha'\tau},
\nonumber \\
\label{eq:H_multiorbital}
\end{eqnarray}
where $i,j$ are site indices within the 1D chain, 
$\tau,\tau'$ are spin indices, 
and $\alpha,\beta = 1$ ($d_{xy}$) and $2$ ($d_{yz}$) are orbital indices. 
The first term describes the electron hopping, and 
the second term denotes the onsite Coulomb interactions 
where we use the standard parametrizations, 
\begin{eqnarray}
U_{\alpha\beta,\alpha'\beta'} 
&=& U' \delta_{\alpha\alpha'} \delta_{\beta\beta'} 
+ J_{\rm H} (\delta_{\alpha\beta'} \delta_{\beta\alpha'} +
\delta_{\alpha\beta} \delta_{\alpha'\beta'}),
\label{eq:defU}
\\
U &=& U_{11,11} = U_{22,22} = U' + 2J_{\rm H}.
\label{eq:relationUJ}
\end{eqnarray}
Here, we neglect the relativistic spin-orbit coupling. 

The perturbation calculation is performed 
in the strong correlation limit $t_{ij}^{\alpha\beta} \ll U, U', J_{\rm H}$ 
by taking an atomic eigenstate with one electron at each site. 
In the edge-sharing configuration as shown in Fig.~\ref{fig:structure} (b), 
the most relevant contribution 
in the transfer integrals $t_{ij}^{\alpha\beta}$ 
is the overlap between the nearest-neighbor (NN) pairs
with the same orbitals lying in the same plane, 
which is called the $\sigma$ bond. 
The $\sigma$-bond transfer integrals are $t_{i,i+1}^{11}$ for NN pairs 
in the $xy$ plane and $t_{i,i+1}^{22}$ for those in the $yz$ plane 
as shown in Fig.~\ref{fig:structure} (b). 
These two types of the transfer integrals take the same value, and 
we denote them by $t_{\sigma}$ in the following. 
Other transfer integrals are much smaller than $t_{\sigma}$; 
in particular, $t_{i,i+1}^{12} = t_{i,i+1}^{21} = 0$ between NN sites 
from the symmetry. 
In the present study, we take account of the $\sigma$-bond contributions only, 
and neglect other transfer integrals.\cite{effectofJT} 
The approximation is known to give reasonable results 
for spinel oxides which also have edge-sharing network of octahedra.
\cite{Tsunetsugu2003} 

The second order perturbation in $t_{\sigma}$ gives 
the effective spin-orbital Hamiltonian in the form 
\begin{eqnarray}
\mathcal{H}_{\rm so} &=& -J \sum_{i} \ 
(h_{i,i+1}^{\rm oAF} + h_{i,i+1}^{\rm oF}), 
\label{eq:H_so}
\\
h_{i,i+1}^{\rm oAF} &=& (A + B \mib{S}_i \cdot \mib{S}_{i+1})
\left(\frac{1}{2} - 2T_i T_{i+1}\right),
\label{eq:h^oAF in T}
\\
h_{i,i+1}^{\rm oF} &=& 
C \left(\frac{1}{4} - \mib{S}_i \cdot \mib{S}_{i+1}\right) 
\left(\frac{1}{2} \pm T_i\right)\left(\frac{1}{2} \pm T_{i+1}\right), 
\nonumber \\
\label{eq:h^oF in T}
\end{eqnarray}
where 
$\mib{S}_i$ is the $S=1/2$ spin operator at site $i$ and 
$T_i$ is the Ising isospin which describes the orbital state at site $i$ 
as $T_i = +1/2 (-1/2)$ when the $d_{xy}$ ($d_{yz}$) orbital is occupied. 
Note that the $\sigma$-bond transfer integral $t_\sigma$, 
which is orbital diagonal and does not mix different orbitals, 
leads to the Ising nature of the orbital isospin interaction. 
The signs in Eq.~(\ref{eq:h^oF in T}) take $+$ ($-$) 
for the bonds within the $xy$ ($yz$) plane. 
The coupling constants in Eqs.~(\ref{eq:H_so})-(\ref{eq:h^oF in T}) 
are given by parameters in Eq.~(\ref{eq:H_multiorbital}) as 
\begin{eqnarray}
J &=& \frac{(t_{\sigma})^2}{U}, 
\label{eq:J}
\\
A &=& \frac{3}{4(1-3\eta)} + \frac{1}{4(1-\eta)}, 
\\
B &=& \frac{1}{1-3\eta} - \frac{1}{1-\eta}, 
\\
C &=& \frac43 \left( \frac{1}{1+\eta} + \frac{2}{1-\eta} \right),
\label{eq:couplingC}
\\
\eta &=& \frac{J_{\rm H}}{U}, 
\label{eq:defeta}
\end{eqnarray} 
where we use Eq. (\ref{eq:relationUJ}). 
The realistic value of $\eta$ will be estimated as $\eta \sim 0.1$ later. 
Therefore, we consider that $A$, $B$, and $C$ are all positive 
in the following. 

We note that a part of the spin-orbital interactions, i.e., 
$h_{i,i+1}^{\rm oF}$ in Eq.~(\ref{eq:h^oF in T}) 
takes a similar form to the model proposed 
in Ref.~\onlinecite{Konstantinovic2004}. 
Our model derived from the multiorbital Hubbard model 
contains both ferro and antiferro types of orbital interactions 
as well as the nontrivial contributions
which are missed in the previous model 
in Ref.~\onlinecite{Konstantinovic2004}. 
We will show that these factors play important roles in the thermodynamics. 
We also note that a spin-orbital model similar to Eq.~(\ref{eq:H_so}) 
was studied in Ref.~\onlinecite{Katoh1999}. 
The model was derived for the corner-sharing configuration of the octahedra, 
while our model is for the edge-sharing configuration. 

The orbital-lattice coupling term $\mathcal{H}_{\rm JT}$ in Eq.~(\ref{eq:H}) 
is given in the form 
\begin{equation}
\mathcal{H}_{\rm JT} = \gamma \sum_i Q_i T_i + \frac{1}{2}\sum_i Q_i^2.
\label{eq:H_JT}
\end{equation}
The first term describes the JT coupling 
where $\gamma$ is the electron-lattice coupling constant and 
$Q_i$ is the amplitude of the JT distortion 
which couples to the remaining two-fold degeneracy of 
the $d_{xy}$ and $d_{yz}$ orbitals.
The second term denotes the elastic energy of the JT distortion. 
For simplicity, we neglect the kinetic energy of phonons and 
regard $Q_i$ as a classical variable. 
Here, we note that besides the onsite term 
there may be intersite 
interactions of JT distortions such as $\sum_{ij} Q_i Q_j$.
However, in the self-consistent scheme described in Sec.~\ref{sec:method}, 
which we will employ in the present numerical study, 
the intersite interactions 
do not affect the results except for a shift of the critical temperature. 
Such effect can be renormalized into the $\gamma$ term in Eq.~(\ref{eq:H_JT}), 
and therefore, we do not explicitly include the intersite term 
in the present Hamiltonian. 

In addition to the above two terms $\mathcal{H}_{\rm so}$ and 
$\mathcal{H}_{\rm JT}$, 
we also consider the interchain coupling term $\mathcal{H}_\perp$ 
in Eq.~(\ref{eq:H}).
We may consider two contributions to $\mathcal{H}_\perp$. 
One is the spin-orbital exchange interaction arising from 
the interchain transfer integrals of electrons, and the other is 
the interchain interaction of JT distortions.
The former is expected to be negligibly small 
due to a rapid decay and a large spatial anisotropy of the $t_{2g}$ wave functions. 
We therefore ignore it and 
take account of the latter JT contribution only.
The explicit form of the JT coupling depends on the details of the  
lattice structure among 1D chains, and may be very complicated. 
In the present study, we assume the following simple form 
\begin{equation}
\mathcal{H}_\perp = \lambda \sum_{\langle i,j \rangle} Q_i Q_j, 
\label{eq:H_perp}
\end{equation}
where the summation is taken over the NN sites 
in four neighboring chains. 
Since the TiO$_6$ chains are well separated 
by SiO$_4$ tetrahedra in the pyroxene compounds, 
the coupling constant $\lambda$ is expected to be small. 
In the following numerical calculations, 
we will treat the interchain JT coupling in a mean-field approximation, 
which is well justified for weakly-coupled 1D systems. 
\cite{Imry1975}

Finally, we discuss the realistic values of the parameters. 
The $\sigma$-bond transfer integral is estimated 
as $t_{\sigma} \sim -0.3$eV for pyroxene vanadium oxides. 
\cite{Millet1999}
Typical estimates for Coulomb interaction parameters are 
$U \sim 5-6$eV and $J_{\rm H} \sim 0.6-0.7$eV. 
\cite{Mizokawa1996}
Hence, $\eta$ in Eq.~(\ref{eq:defeta}) 
is a small parameter of the order of 0.1. 
From the estimates of $t_{\sigma}$ and $U$, the exchange energy scale $J$ 
in Eq.~(\ref{eq:J}) is estimated as $J \sim 200$K. 
As for the JT couplings $\gamma$ and $\lambda$, 
since there is no experimental estimates to our knowledge, 
we treat them as parameters 
which are determined by the comparison between the numerical results 
and the experimental data in Sec.~\ref{sec:comparison}.

\section{mean-field analysis}
\label{sec:MF}

Before going into the numerical study of the thermodynamics of 
the model (\ref{eq:H}), 
we apply mean-field arguments to capture 
the behavior in the ground state and in the high-temperature phase. 
For simplicity, 
we consider only the spin-orbital part $\mathcal{H}_{\rm so}$ in this section. 

\subsection{Ground state}
\label{sec:gs}

\begin{figure}
\includegraphics[width=70mm]{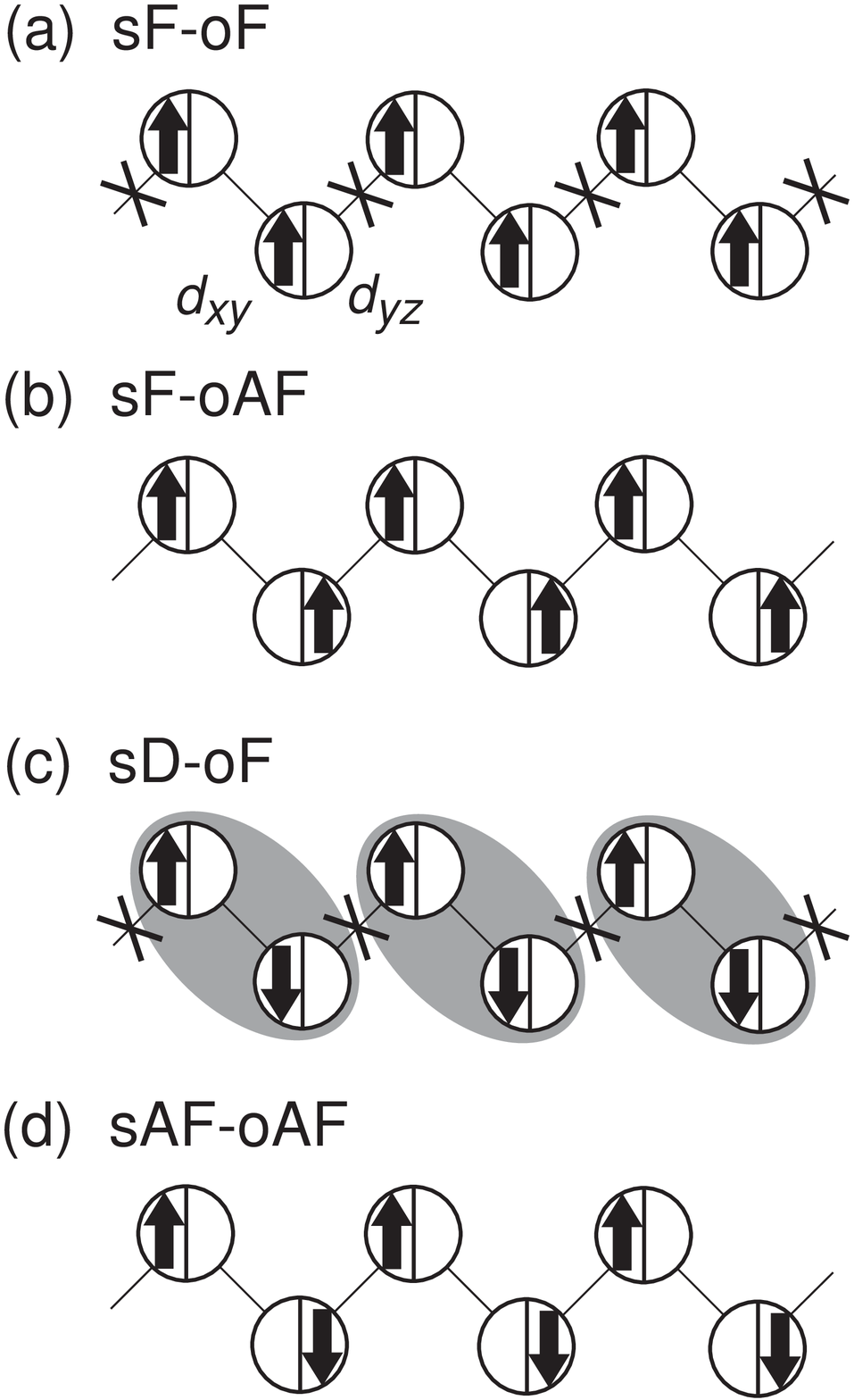}
\caption{
Spin and orbital ordering patterns in  
(a) the spin-ferro and orbital-ferro (sF-oF), 
(b) the spin-ferro and orbital-antiferro (sF-oAF), 
(c) the spin-dimer and orbital-ferro (sD-oF), and 
(d) the spin-antiferro and orbital-antiferro (sAF-oAF) states. 
Circles denote the lattice sites, and 
two separated spaces inside them denote two different orbital states 
$d_{xy}$ and $d_{yz}$ as indicated in (a). 
Arrows denote spins, and gray ovals in (c) represent 
the spin-singlet pairs. 
Crosses in (a) and (c) denote the disconnected bonds 
by the ferro-type orbital ordering. 
See the text for details. 
} 
\label{fig:SOpattern}
\end{figure}

In the ground state, we here consider four different types of ordered states 
as shown in Fig.~\ref{fig:SOpattern} schematically; 
(a) the spin-ferro and orbital-ferro (sF-oF), 
(b) the spin-ferro and orbital-antiferro (sF-oAF), 
(c) the spin-antiferro and orbital-ferro, and 
(d) the spin-antiferro and orbital-antiferro (sAF-oAF) states. 
For simplicity, we consider the fully polarized states 
for the spin-ferro, the orbital-ferro, and the orbital-antiferro orderings, 
where quantum fluctuations do not play a crucial role. 
(Note that the orbital isospin is the Ising spin in the present model.)
The important point is that as easily shown 
by Eqs.~(\ref{eq:h^oAF in T}) and (\ref{eq:h^oF in T}), 
the orbital-ferro ordering, i.e., 
$\langle T_i T_{i+1} \rangle = 1/4$ and 
$\langle T_i \rangle = 1/2$ (or $-1/2$) for all $i$, 
disconnects every other bonds. 
Hence, when the spin coupling is antiferromagnetic, 
the orbital-ferro ordering leads to the spin-singlet formation, i.e., 
$\langle \mib{S}_i \cdot \mib{S}_{i+1} \rangle = -3/4$ 
for remaining isolated NN pairs. 
Therefore, the ordered state (c) is denoted as 
the spin-dimer and orbital-ferro (sD-oF) state. 
We note that a similar mechanism of the singlet formation driven by 
orbital ordering has been proposed 
for a related model.\cite{Katoh1999}

The ground-state energy for each ordered state is calculated 
by replacing the spin and orbital-isospin operators in Eq.~(\ref{eq:H_so}) 
by the following expectation values; 
\begin{eqnarray}
\langle \mib{S}_i \cdot \mib{S}_{i+1} \rangle 
= \frac14 && \mbox{for sF pairs}, 
\\
\langle \mib{S}_i \cdot \mib{S}_{i+1} \rangle 
= -\frac34 && \mbox{for sD pairs}, 
\\
\langle \mib{S}_i \cdot \mib{S}_{i+1} \rangle 
= -s && \mbox{for sAF pairs}, 
\end{eqnarray}
where $s$ is a positive parameter less than $3/4$ 
(we do not need the precise value of $s$), 
and 
\begin{eqnarray}
\langle T_i T_{i+1} \rangle = \frac14, \ \langle T_i \rangle = \pm \frac12
&& \mbox{for oF pairs}, 
\\ 
\langle T_i T_{i+1} \rangle = -\frac14, 
\ \langle T_i \rangle = \pm \frac{(-1)^i}{2} 
&& \mbox{for oAF pairs},
\end{eqnarray}
respectively. 
The obtained ground-state energies per site for the states (a)-(d) are 
\begin{eqnarray}
&& E_{\rm sF-oF} = 0,
\label{eq:E_sF-oF}
\\
&& E_{\rm sF-oAF} = -J \left( A + \frac{B}{4} \right), 
\label{eq:E_sF-oAF}
\\
&& E_{\rm sD-oF} = -J \frac{C}{2},
\label{eq:E_sD-oF}
\\
&& E_{\rm sAF-oAF} = -J (A - s B), 
\label{eq:E_sAF-oAF}
\end{eqnarray}
respectively. 

We compare Eqs.~(\ref{eq:E_sF-oF})-(\ref{eq:E_sAF-oAF}) 
by using Eqs.~(\ref{eq:J})-(\ref{eq:couplingC}) 
and obtain 
\begin{equation}
(E_{\rm sF-oAF} \ \mbox{or} \ E_{\rm sD-oF}) <
E_{\rm sAF-oAF} < E_{\rm sF-oF} . 
\end{equation}
Hence, the ground state is either 
(b) sF-oAF or (c) sD-oF state. 
From the equation of $E_{\rm sF-oAF} = E_{\rm sD-oF}$,
we obtain the critical value of $\eta$ 
for the transition between the two phases as 
\begin{equation}
\eta_{\rm c} = \frac13 (\sqrt{73}-8) \simeq 0.18.  
\end{equation}
Namely, we have the ground-state phase transition 
between the sD-oF and sF-oAF phases by changing $\eta$; 
the sD-oF (sF-oAF) phase is stable for $\eta < \eta_{\rm c}$ 
($\eta > \eta_{\rm c}$). 
Hence, for a realistic value of $\eta \sim 0.1$, 
the ground state is predicted to be the sD-oF phase 
within the mean-field argument. 
This will be confirmed by the numerical calculations 
in Sec.~\ref{sec:results}. 
We will also show that the sF-oAF state is realized 
for $\eta > \eta_{\rm c}$ in Appendix B. 

The $\eta$-controlled phase transition 
is understood by the competition between the spin superexchange interaction 
and the Hund's-rule coupling. 
The former comes from the perturbation process within the same orbitals 
and favors the spin-singlet state. 
\cite{Anderson1959}
The latter enhances the energy gain from 
the perturbation process between different orbitals 
and stabilizes the sF-oAF state. 
It is known that the sF-oAF state is favored 
by a finite Hund's-rule coupling ($\eta > 0$) 
in multiorbital systems with transfer integrals
$t^{11} = t^{22} \neq 0$ between all the NN sites. 
\cite{Roth1966,Inagaki1973}
In the present systems, 
the specific form of the transfer integrals 
due to the zig-zag lattice structure gives a chance 
to stabilize the sD-oF state in the small $\eta$ regime.

\subsection{High-temperature para phase}
\label{sec:high-T}

At high temperatures, both spin and orbital are disordered. 
We here examine spin and orbital correlations in the para phase 
by a mean-field type argument. 

First, to focus on the spin degree of freedom, 
we replace the orbital isospin operators 
in the effective model (\ref{eq:H_so}) with their mean values, i.e., 
$T_i \to \langle T_i \rangle = 0$ and 
$T_i T_{i+1} \to \langle T_i T_{i+1} \rangle$. 
The resultant effective spin Hamiltonian reads 
\begin{equation}
\mathcal{H}_{\rm s}^{\rm MF} 
= \sum_{i} 
\left( J_{i,i+1}^{\rm s} \mib{S}_i \cdot \mib{S}_{i+1} 
     - K_{i,i+1}^{\rm s} \right),
\label{eq:H_MF^s}
\end{equation}
where 
\begin{eqnarray}
J_{i,i+1}^{\rm s} &=& J \left[ 
\frac14 (C-2B)
+ (C + 2B) \langle T_i T_{i+1} \rangle \right], 
\label{eq:JsT}
\\
K_{i,i+1}^{\rm s} &=& \frac{J}{4} \left[
\frac14 (C + 8A) + (C - 8A) \langle T_i T_{i+1} \rangle
\right].
\label{eq:KsT}
\end{eqnarray}
In the high-temperature limit, $\langle T_i T_{i+1} \rangle$ becomes zero 
and the effective spin exchange constant becomes 
\begin{equation}
J_{i,i+1}^{\rm s} (T \to \infty) = \frac{J}{4}(C-2B), 
\end{equation}
which is positive for $\eta \sim 0.1$ 
and independent of $i$. 
Hence, we end up with a 1D AF spin Heisenberg model.

On the other hand, we can also consider a reduced Hamiltonian 
for the orbital degree of freedom by replacing the spin operators 
in the model (\ref{eq:H_so}) with their mean values. 
The result is 
\begin{equation}
\mathcal{H}_{\rm o}^{\rm MF} 
= \sum_{i} 
\left[ J_{i,i+1}^{\rm o} T_i T_{i+1} - L_{i,i+1}^{\rm o} (T_i + T_{i+1}) 
- K_{i,i+1}^{\rm o} \right],
\label{eq:H_MF^o}
\end{equation}
where 
\begin{eqnarray}
J_{i,i+1}^{\rm o} &=& J \left[ \left( 2A - \frac{C}{4} \right) 
+ (C + 2B) \langle \mib{S}_i \cdot \mib{S}_{i+1} \rangle \right], 
\nonumber \\
\label{eq:JoT}
\\
L_{i,i+1}^{\rm o} &=& (-1)^i J \frac{C}{2} \left(
\frac14 - \langle \mib{S}_i \cdot \mib{S}_{i+1} \rangle 
\right), 
\label{eq:LoT}
\\
K_{i,i+1}^{\rm o} &=& \frac{J}{4} \left[
\left( 2A + \frac{C}{4} \right) - 
(C - 2B) \langle \mib{S}_i \cdot \mib{S}_{i+1} \rangle
\right]. 
\nonumber \\
\label{eq:KoT}
\end{eqnarray}
In the high-temperature limit, 
$\langle \mib{S}_i \cdot \mib{S}_{i+1} \rangle = 0$ 
and the effective isospin coupling constant becomes
\begin{equation}
J_{i,i+1}^{\rm o}(T \to \infty) = J \left( 2A - \frac{C}{4} \right). 
\end{equation}
This coupling constant is also positive 
for $\eta \sim 0.1$, and therefore, 
we obtain a 1D antiferro-type Ising isospin model. 
[Note that the second term in Eq.~(\ref{eq:H_MF^o}) cancels out 
when $\langle \mib{S}_i \cdot \mib{S}_{i+1} \rangle = 0$.]

The above arguments indicate that 
both spin and orbital correlations are antiferro type 
in the high-temperature limit in our effective model (\ref{eq:H}). 
On the contrary, as discussed in Sec.~\ref{sec:gs}, 
the ground state is either the sF-oAF or sD-oF state.
This suggests that either spin or orbital correlation has to change 
from antiferro to ferro type as decreasing temperature. 
It is implied by Eq.~(\ref{eq:JsT}) that 
development of antiferro-type orbital correlations 
$\langle T_i T_{i+1} \rangle < 0$ may change the sign of $J_{i,i+1}^{\rm s}$ 
from positive (antiferro) to negative (ferro). 
In a similar way, Eq.~(\ref{eq:JoT}) suggests that 
development of AF spin correlations 
$\langle \mib{S}_i \cdot \mib{S}_{i+1} \rangle < 0$ may induce 
the sign change of $J_{i,i+1}^{\rm o}$. 
In this manner, the spin and orbital correlations compete with each other.
The mean-field-level argument is clearly insufficient to clarify 
the competition, and we will employ the more powerful numerical analysis 
in the next section.

\section{Numerical Analysis}
\label{sec:results}

\subsection{Method}
\label{sec:method}

To study thermodynamic properties of the model (\ref{eq:H}),
we apply the quantum transfer matrix (QTM) method\cite{Betsuyaku1984} 
combined with the mean-field treatment of 
the JT distortions.
Here, we describe the scheme of our analysis.

Since the present model (\ref{eq:H}) is highly 1D anisotropic, 
it is justified to treat the weak interchain coupling $\mathcal{H}_\perp$ 
as the mean field\cite{Imry1975} in the form
\begin{equation}
\tilde{\mathcal{H}}_\perp
= \lambda \sum_{\langle i,j \rangle} Q_i \tilde{Q}_j
= - z |\lambda| \sum_i Q_i^2, 
\end{equation}
where $\tilde{Q}_j$ is the mean-field value of the JT distortion 
at site $j$ and $z$ is the number of NN chains, i.e., 
$z=4$ in the present materials. 
As a result, the total Hamiltonian, 
$\mathcal{H}_{\rm so} + \mathcal{H}_{\rm JT} + \tilde{\mathcal{H}}_\perp$, 
is reduced to an effective 1D spin-orbital-lattice coupled model 
in the form
\begin{eqnarray}
\tilde{\mathcal{H}} &=& 
\mathcal{H}_{\rm so} + \gamma \sum_i Q_i T_i 
+ \frac{1-2z|\lambda|}{2}\sum_i Q_i^2
\nonumber \\
&=& 
\mathcal{H}_{\rm so} + \bar{\gamma} \sum_i \bar{Q}_i T_i 
+ \frac{1}{2}\sum_i \bar{Q}_i^2,
\label{eq:tildeH}
\end{eqnarray}
where the JT coupling and distortion are rescaled as 
$\bar{\gamma} = \gamma/\sqrt{1-2z|\lambda|}$ 
and $\bar{Q}_i = \sqrt{1-2z|\lambda|}Q_i$, respectively.
For later use, we define the JT stabilization energy as 
\begin{equation}
\Delta_{\rm JT} = \bar{\gamma}^2/8,
\label{eq:D_JT}
\end{equation}
which is the energy gain by the JT distortion 
in the absence of $\mathcal{H}_{\rm so}$.

The optimal values of the JT distortions $\{ \bar{Q}_i \}$ are determined 
in a self-consistent way.
We start from an initial guess of $\{ \bar{Q}_i \}$ and 
calculate the expectation values of the isospin $\langle T_i \rangle$ 
by applying the QTM method to the effective 1D model $\tilde{\mathcal{H}}$. 
Note that the QTM calculation is numerically exact and 
includes all the fluctuations in spin and orbital degrees of freedom 
for a given set of $\{ \bar{Q}_i \}$. 
See Appendix A for the details of the QTM method. 
The obtained values of $\langle T_i \rangle$ are used 
to determine $\{ \bar{Q}_i \}$ self-consistently. 
The self-consistent equation is obtained by the energy minimization 
$\delta \langle \mathcal{H} \rangle / \delta \bar{Q}_i = 0$ which gives 
\begin{equation}
\bar{Q}_i^{\rm new} = - \bar{\gamma} \langle T_i \rangle.
\end{equation}
The values of $\{ \bar{Q}_i^{\rm new} \}$ are used as inputs 
for the next QTM calculation.
We iterate the procedure until $\{ \bar{Q}_i \}$ converge 
to optimal values. 

Thermodynamic properties are obtained 
for the optimal values of $\{ \bar{Q}_i \}$. 
We calculate the magnetization per site,
\begin{equation}
m = \frac{1}{L} \sum_i \langle S^z_i \rangle,
\label{eq:Sz}
\end{equation}
and the average values of NN spin and orbital correlations, 
\begin{eqnarray}
C_{\rm s} &=& \frac{1}{L} 
\sum_i \langle \mib{S}_i \cdot \mib{S}_{i+1} \rangle,
\label{eq:Cspin}
\\
C_{\rm o} &=& \frac{1}{L} \sum_i \langle T_i T_{i+1} \rangle,
\label{eq:Corb}
\end{eqnarray}
respectively, where $L$ is the length of the chain.
We also define the NN spin correlations on odd and even bonds as
\begin{eqnarray}
C_{\rm s}^{\rm odd} &=& \frac{2}{L} \sum_{i \in {\rm odd}}
\langle \mib{S}_i \cdot \mib{S}_{i+1} \rangle,
\label{eq:Cspin_odd}
\\
C_{\rm s}^{\rm even} &=& \frac{2}{L} \sum_{i \in {\rm even}}
\langle \mib{S}_i \cdot \mib{S}_{i+1} \rangle,
\label{eq:Cspin_even}
\end{eqnarray}
respectively, to detect the spin dimerization.
See Eqs.~(\ref{eq:m in App})-(\ref{eq:C_s_even in App}) in Appendix A 
for the calculation of these quantities.
Besides, to calculate the magnetic susceptibility $\chi$, 
we perform the calculation 
within the same framework 
for the system under the external magnetic field 
whose Hamiltonian reads
\begin{equation}
\tilde{\mathcal{H}} - g \mu_B H \sum_i S_i^z,
\end{equation}
where $g$ is the $g$-factor and $\mu_B$ is the Bohr magneton, 
and $H$ is the external magnetic field.
The susceptibility is obtained from the numerical derivative of $m$ as 
\begin{equation}
\chi = \frac{m(\Delta H)-m(0)}{\Delta H},
\label{eq:chi}
\end{equation}
where we use $\Delta H = 0.001J/(g\mu_B)$.

The QTM method allows us to treat the system 
in the thermodynamic limit $L \to \infty$ directly (see Appendix A). 
Instead, the results contain 
a systematic error due to a finite Trotter number $M$. 
We need to check carefully the convergence of the results 
with increasing $M$.
The values of $M$ used in the following calculations 
are up to $M = 4$ for the case of $\bar{\gamma}=0$ 
while up to $M = 3$ for $\bar{\gamma} > 0$.
Although these values of $M$ appear to be rather small, 
we will show that the $M$-convergence is excellent
for the temperature range 
in the present study, $T \simge 0.1J$.

\subsection{Results}
\label{sec:numerical results}

In this section, we show our numerical results 
for thermodynamic properties of the model (\ref{eq:H}) 
obtained by the method in the previous section.
In Sec.~\ref{sec:interplay}, we clarify nontrivial feedback effects 
between orbital and spin degrees of freedom 
in the absence of the JT coupling. 
In Sec.~\ref{sec:comparison}, we switch on the JT coupling, and 
compare our numerical results with the experimental data.

\subsubsection{Interplay between orbital and spin}
\label{sec:interplay}

\begin{figure}
\includegraphics[width=80mm]{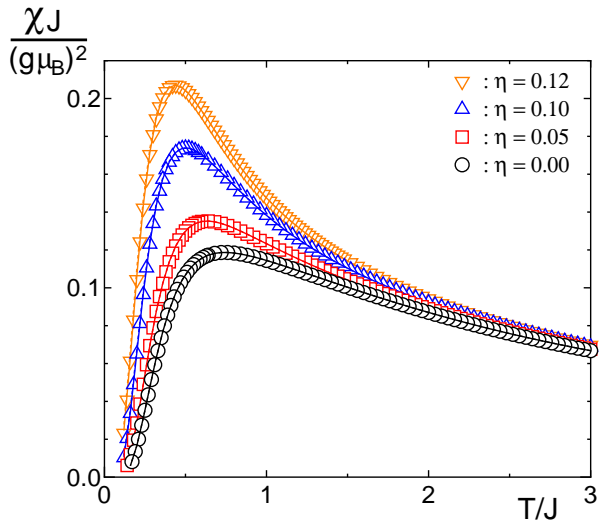}
\caption{
Temperature dependence of the magnetic susceptibility $\chi$ 
for $\eta = 0.0, 0.05, 0.10$, and $0.12$. 
Symbols represent the results for $M = 4$ while 
solid and dotted curves are those for $M = 3$ and $M = 2$, respectively.
} 
\label{fig:chi-eta}
\end{figure}

\begin{figure}
\includegraphics[width=80mm]{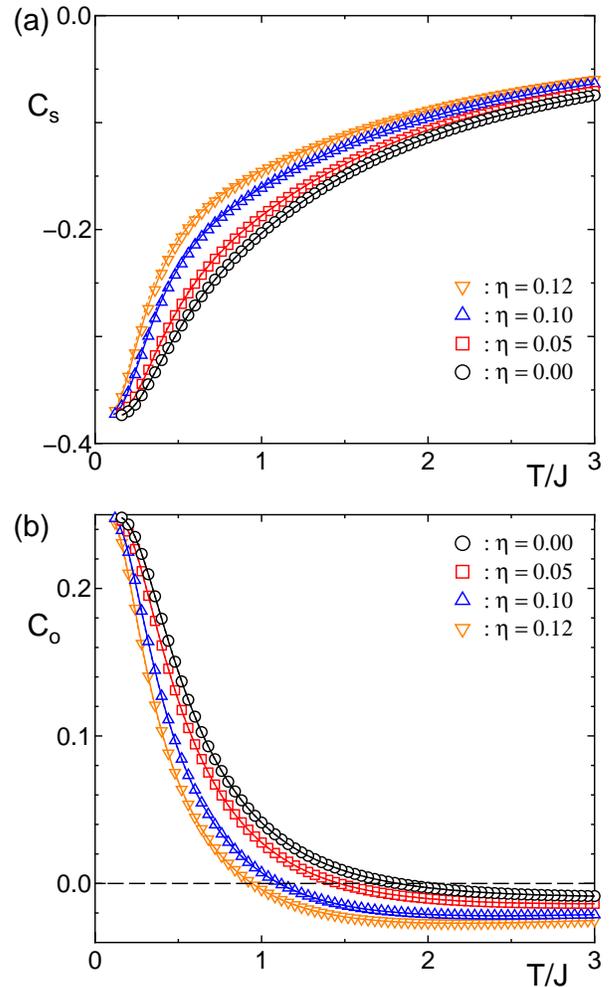}
\caption{
Temperature dependences of 
(a) the nearest-neighbor spin correlation $C_{\rm s}$ 
and (b) the nearest-neighbor orbital correlation $C_{\rm o}$
for $\eta = 0.0, 0.05, 0.10$, and $0.12$.
Symbols represent the results for $M = 4$ while 
solid and dotted curves are those for $M = 3$ and $M = 2$, respectively.
} 
\label{fig:cor-eta}
\end{figure}

\begin{figure}
\includegraphics[width=80mm]{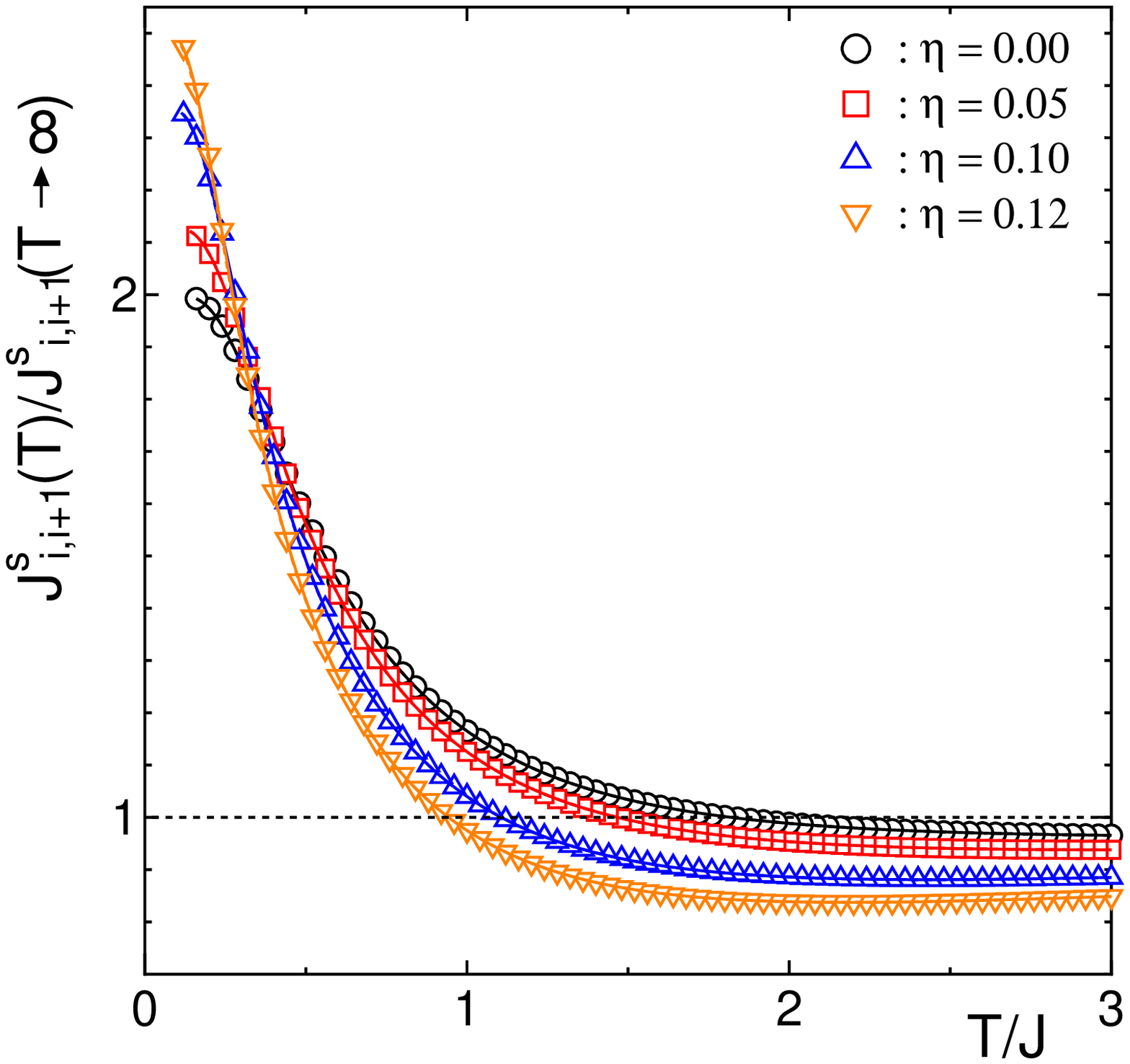}
\caption{
Effective spin exchange coupling $J_{i,i+1}^{\rm s}$ in Eq.~(\ref{eq:JsT}) 
for $\eta = 0.0, 0.05, 0.10$, and $0.12$.
The results are scaled by their values in the high-temperature limit, 
$J_{i,i+1}^{\rm s}(T \to \infty)$.
} 
\label{fig:JsT}
\end{figure}

First, we show our results in the absence of the JT coupling 
($\bar{\gamma} = 0$).
In this case, the model is reduced to the 1D spin-orbital model 
$\mathcal{H}_{\rm so}$ [Eq.~(\ref{eq:H_so})], and 
we can obtain thermodynamic properties 
by performing the QTM calculation without any self-consistent iteration. 

Figure~\ref{fig:chi-eta} shows the results of the magnetic susceptibility 
$\chi$ in Eq.~(\ref{eq:chi}) for several values of $\eta$.
Note that the results for different values of 
the Trotter number $M = 2, 3$, and $4$ 
almost coincide with each other
as shown in the figure, 
which ensures the convergence of the QTM results 
in this parameter range. 
The magnetic susceptibility at high temperatures increases 
as temperature decreases, and exhibits a broad peak at some temperature. 
At lower temperatures, 
it decays rapidly and approaches zero as $T \to 0$.
While the peak becomes sharper and shifts to a lower temperature region 
as $\eta$ increases,
the rapid decay of $\chi$ at low temperatures is commonly observed. 
This indicates that our spin-orbital model $\mathcal{H}_{\rm so}$ 
in Eq.~(\ref{eq:H_so}) exhibits 
the spin-singlet ground state in the small $\eta$ regime 
as predicted in Sec.~\ref{sec:gs}. 
(See Appendix B for the larger $\eta$ regime.)

To clarify the nature of the system in more detail, 
in Fig.~\ref{fig:cor-eta}, we show the results of 
NN spin and orbital correlations $C_{\rm s}$ and $C_{\rm o}$ 
defined in Eqs.~(\ref{eq:Cspin}) and (\ref{eq:Corb}), respectively.
The $M$-dependence of the results is negligible also for these quantities. 
We note that at finite temperatures no true long-range order appears 
in the 1D spin-orbital model $\mathcal{H}_{\rm so}$, 
and hence the NN spin correlations are uniform, i.e., 
$C_{\rm s} = C_{\rm s}^{\rm odd} = C_{\rm s}^{\rm even}$.
As $T \to 0$, the spin and orbital correlations converge to 
$C_{\rm s} = -3/8$ and $C_{\rm o} = 1/4$, respectively. 
These are indeed the values expected in the sD-oF ground state 
shown in Fig.~\ref{fig:SOpattern} (c), where 
$\langle \mib{S}_i \cdot \mib{S}_{i+1} \rangle$ takes the value of 
$-3/4$ or $0$ alternatively from bond to bond 
while $\langle T_i^z T_{i+1}^z \rangle = 1/4$ for all bonds.
Hence, we conclude that the ground state of the system is the sD-oF state 
in Fig.~\ref{fig:SOpattern} (c) 
for the realistic values of $\eta \sim 0.1$, 
and the rapid decay of $\chi$ at low temperatures in Fig.~\ref{fig:chi-eta} 
is due to the spin-singlet formation in the ground state. 

Let us examine 
the finite-temperature behavior of $C_{\rm s}$ and $C_{\rm o}$ 
in Fig.~\ref{fig:cor-eta}. 
At high temperatures, both the spin and orbital correlations 
are negative, i.e., antiferro type, being consistent 
with the mean-field prediction in Sec.~\ref{sec:high-T}.
Here, the spin and orbital antiferro correlations 
compete with each other 
since the spin-antiferro correlation favors the orbital-ferro one 
and vice versa 
as indicated in the form of $\mathcal{H}_{\rm so}$ in Eq.~(\ref{eq:H_so}).
We find that in this range of $\eta$ 
the spin-antiferro correlation grows 
more rapidly than the orbital correlation as $T$ decreases.
This growth of $C_{\rm s}$ suppresses $C_{\rm o}$ and
causes the sign change of $C_{\rm o}$ from antiferro to ferro type.
Once $C_{\rm o}$ becomes ferro type, $C_{\rm s}$ and $C_{\rm o}$ develop 
cooperatively, and eventually 
the ferro-type orbital correlation induces the spin-dimer formation 
in the ground state. 
These results clearly indicate that our model $\mathcal{H}_{\rm so}$
shows a nontrivial feedback effect 
between orbital and spin degrees of freedom at finite temperatures.

The strong interplay between orbital and spin 
also shows up in the temperature dependence of 
the magnetic susceptibility at high temperatures. 
There, we expect a deviation from the Curie-Weiss behavior 
due to the interplay 
since $J_{i,i+1}^{\rm s}$ depends on 
the NN orbital correlation $\langle T_i T_{i+1} \rangle$ 
as mentioned in Sec.~\ref{sec:high-T}. 
The deviation can be monitored by 
the temperature dependence of the effective spin exchange coupling 
$J_{i,i+1}^{\rm s}$ in Eq.~(\ref{eq:JsT}), 
since it gives 
an effective Curie-Weiss temperature $\Theta_{\rm CW}$. 
(In the mean-field approximation, $\Theta_{\rm CW} = J_{i,i+1}^{\rm s}/2$ 
in the 1D spin-$1/2$ system with NN interaction.) 

In Fig.~\ref{fig:JsT}, we show the temperature dependence of 
$J_{i,i+1}^{\rm s}$ 
in Eq.~(\ref{eq:JsT}) 
scaled by the value in the high-temperature limit.
The results clearly show that $J_{i,i+1}^{\rm s}$ 
is temperature-dependent even in 
the temperature range where the Curie-Weiss fitting to 
the experimental data has been attempted in Ref.~\onlinecite{Isobe2002}. 
(The energy scale of $J$ will be estimated as $200-300$K 
from the fitting in the next section.)
We need further careful considerations 
in the fitting of the experimental data of the magnetic susceptibility 
to estimate the model parameters.

\subsubsection{Comparison with experimental results}
\label{sec:comparison}

\begin{figure}
\includegraphics[width=80mm]{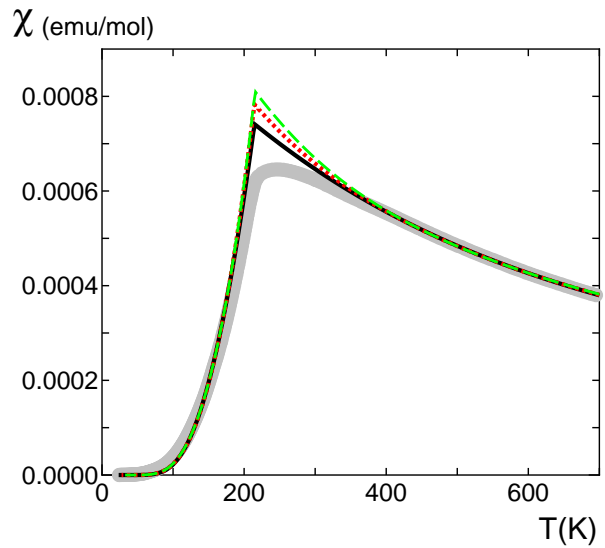}
\caption{
Comparison between numerical and experimental results 
of the magnetic susceptibility $\chi$.
Solid, dotted, and dashed curves represent the numerical results 
for $(J, \eta, g, \Delta_{\rm JT}) = (213$K$, 0.05, 1.85, 86$K$)$, 
$(250$K$, 0.10, 1.87, 90$K$)$, and $(296$K$, 0.12, 1.90, 89$K$)$, 
respectively.
Bold gray curve represents the experimental data 
in Ref. \onlinecite{Isobe2002}.
} 
\label{fig:experiment}
\end{figure}

Next, we show our results in the presence of the JT coupling, i.e., 
for the case of $\bar{\gamma} > 0$ in Eq.~(\ref{eq:tildeH}), and 
compare them with experimental data quantitatively. 
The numerical results shown in the following are obtained from 
the self-consistent iteration of the QTM calculation with $M = 3$, 
and we have confirmed the $M$-convergence 
of the results. 

Figure~\ref{fig:experiment} shows the results of 
the magnetic susceptibility $\chi$.
The results show a singularity at some temperature and 
a sudden drop below there, which corresponds to 
the phase transition to the low-temperature sD-oF phase as discussed later. 
For comparison, we plot the experimental data for NaTiSi$_2$O$_6$,\cite{chi_m}
which increase as $T$ decreases at high temperatures 
and exhibit a sharp drop at $T_{\rm c} = 210$K. 
The numerical results are obtained for several typical 
parameter sets estimated by the following fitting procedure:
In our model, there are four parameters, i.e., 
the effective exchange coupling $J$, 
the ratio of Coulomb interactions $\eta = J_{\rm H}/U$, 
the $g$-factor $g$, and the JT coupling parameter $\bar{\gamma}$.
For a certain value of $\eta$, 
we perform the two-parameter fitting by using $J$ and $g$ 
in the high-temperature regime of $400$K $< T < 700$K, 
where the JT coupling $\bar{\gamma}$ is irrelevant 
in the present scheme of the calculation. 
Using the optimal values of $J$ and $g$, we determine 
$\bar{\gamma}$ to reproduce 
the critical temperature $T_{\rm c} = 210$K.
We thereby obtain the optimal set of $J$, $g$, and $\bar{\gamma}$ 
for a certain value of $\eta$. 

In Fig.~\ref{fig:experiment}, we show the results of the fitting 
for $\eta = 0.05, 0.10$, and $0.12$ as typical examples.
The estimates of the parameters are 
$(J, g, \Delta_{\rm JT}) = (213$K$, 1.85, 86$K$)$, 
$(250$K$, 1.87, 90$K$)$, and $(296$K$, 1.90, 89$K$)$ 
for $\eta = 0.05$, $0.10$, and $0.12$, respectively.
As shown in the figure, the numerical results well 
reproduce the experimental data, 
except for a small deviation near $T_{\rm c}$ (discussed below).
Although we cannot determine the best set of the parameter values 
only from the present fitting, 
we find that the estimates are 
quite reasonable in this $t_{2g}$ compound: 
The estimates of $J \sim 200-300$K are 
comparable to the estimate based on microscopic parameters 
in Sec.~\ref{sec:model}.
As for the $g$-factor, although there is no estimate 
for the present compound as far as we know,
it is known that $g$ becomes slightly smaller than two 
in some $t_{2g}$ compounds, 
\cite{Kataev2003,Yamada1998}
and therefore we believe that our estimates of $g$ are reasonable. 
Furthermore, the estimates of $\Delta_{\rm JT} \sim 90$K, 
which are smaller than the exchange energy $J$,
also appear to be reasonable for $t_{\rm 2g}$ electron systems.
(We will comment on the magnitude of $\Delta_{\rm JT}$ 
in the end of this section.)
Therefore, we believe that our effective model (\ref{eq:H}) 
with realistic parameters can describe successfully 
the peculiar properties of the pyroxene compound NaTiSi$_2$O$_6$.

We note that the small deviation near $T_{\rm c}$ can be attributed to 
the approximation in the treatment of the JT distortion $\bar{Q}_i$:
Our method to determine $\bar{Q}_i$ 
is not able to include effects of the short-range correlations 
and thermal fluctuations of lattice distortions,
which tend to enhance the spin-singlet fluctuation
and suppress the magnetic susceptibility $\chi$.
It is therefore reasonable that 
our method overestimates $\chi$ around the critical point, 
where the effects become significant.
We believe that better agreement near $T_{\rm c}$ may be obtained 
by including such effects, however 
this interesting problem is left for further study.

As shown in Fig.~\ref{fig:experiment}, 
the magnetic susceptibility of our model shows 
an exponential decay at low temperatures well below $T_{\rm c}$. 
We estimate the spin gap $\Delta_{\rm s}$ from 
the results below $0.5 T_{\rm c}$ by the fitting function
\begin{equation}
\chi \propto \exp( -\Delta_{\rm s} / T ).
\end{equation}
The estimates of $\Delta_{\rm s}$ are $665, 640$, and $638$K 
for $\eta = 0.05, 0.10$, and $0.12$, respectively. 
In experiment, the spin gap $\Delta_{\rm s}$ is estimated as 
$\sim 500$K,\cite{Isobe2002} which is comparable to our numerical results. 

\begin{figure}
\includegraphics[width=80mm]{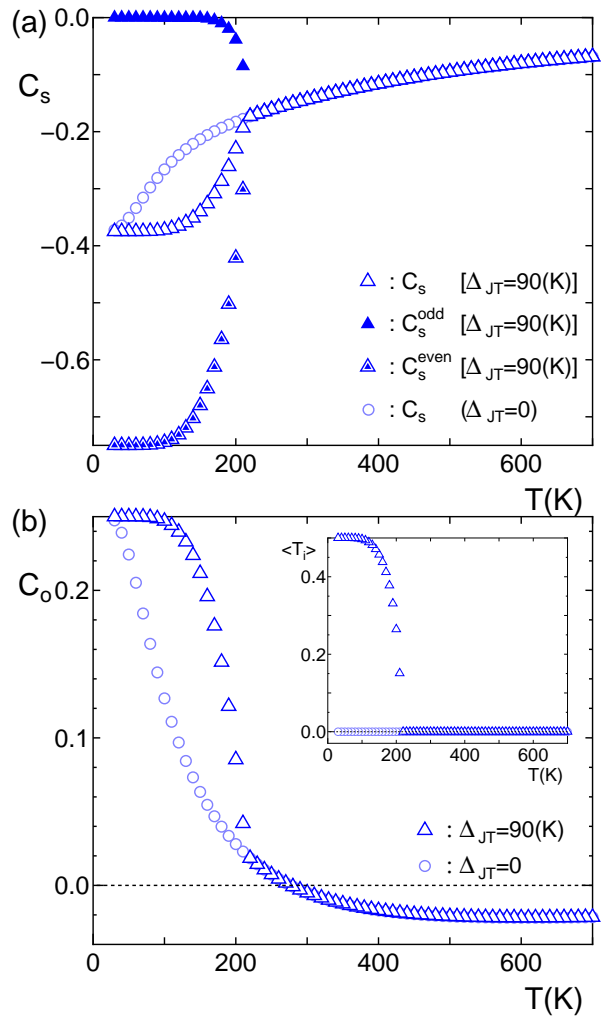}
\caption{
Temperature dependences of 
(a) the nearest-neighbor spin correlations $C_{\rm s}$, 
$C_{\rm s}^{\rm odd}$, and $C_{\rm s}^{\rm even}$, 
and (b) the nearest-neighbor orbital correlation $C_{\rm o}$
for 
$(J, \eta, \Delta_{\rm JT}) = (250$K$, 0.10, 90$K$)$. 
Circles show the data for $\Delta_{\rm JT} = 0$ for comparison. 
Inset in (b): Polarization of orbital isospin.
} 
\label{fig:corJT}
\end{figure}
\begin{figure}
\includegraphics[width=80mm]{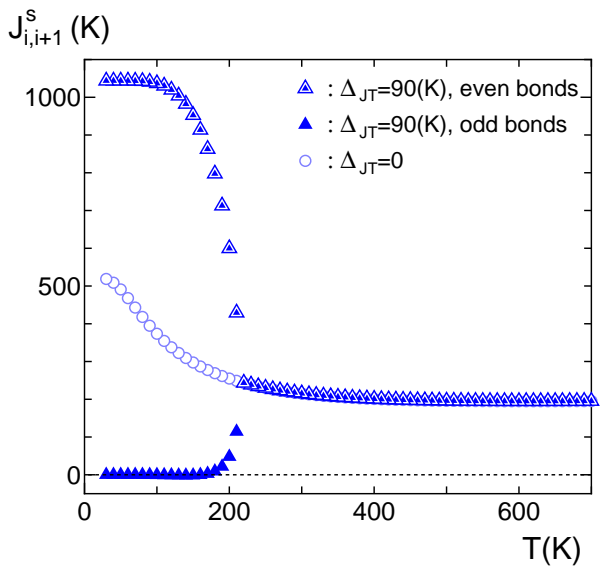}
\caption{
Temperature dependence of the effective spin exchange coupling 
$J_{i,i+1}^{\rm s}$ for 
$(J, \eta, \Delta_{\rm JT}) = (250$K$, 0.10, 90$K$)$. 
Circles show the data for $\Delta_{\rm JT} = 0$ for comparison. 
} 
\label{fig:JsTJT}
\end{figure}

Next, we show the results of the spin and orbital correlations 
in Fig.~\ref{fig:corJT}.
Here, we plot only the results for 
the parameter set of $(J, \eta, \Delta_{\rm JT}) = (250$K$, 0.10, 90$K$)$
as a typical example 
since essentially the same behavior is obtained 
for other parameter sets in Fig.~\ref{fig:experiment}.\cite{noneedg}
The correlations take the same values as those for 
$\Delta_{\rm JT} = 0$ above $T_{\rm c}$. 
Below $T_{\rm c}$, the NN spin correlations on odd and even bonds, 
$C_{\rm s}^{\rm odd}$ and $C_{\rm s}^{\rm even}$, take 
different values and  converge to $-3/4$ or $0$ as $T \to 0$. 
This indicates that the translational symmetry is broken 
in the spin-dimer phase. 
(Note that one of the doubly-degenerate ordered states is selected 
depending on the initial set of $\{ \bar{Q}_i \}$ used 
in the self-consistent calculation.)
On the other hand, the NN orbital correlation $C_{\rm o}$ is uniform, and 
goes to the value of $1/4$ as $T \to 0$. 
As shown in the inset of Fig.~\ref{fig:corJT} (b), 
the isospin polarization $\langle T_i \rangle$ becomes finite 
below $T_{\rm c}$, indicating the long-range orbital ordering. 
These results explicitly show that 
the system below $T_{\rm c}$ is in 
the sD-oF phase in Fig.~\ref{fig:SOpattern} (c).

In Fig.~\ref{fig:JsTJT}, we plot the effective spin exchange coupling 
$J_{i,i+1}^{\rm s}$, corresponding to Eq.~(\ref{eq:JsT}) 
in the case of $\Delta_{\rm JT} = 0$. 
For a finite $\Delta_{\rm JT}$, 
we have to keep the linear terms of $\langle T_i \rangle$ 
in the reduced spin Hamiltonian in Eq.~(\ref{eq:H_MF^s}), 
which give rise to an additional term 
$
- J (-1)^i C/2 \left( \langle T_i \rangle + \langle T_{i+1} \rangle \right)
$
to $J_{i,i+1}^{\rm s}$ in Eq.~(\ref{eq:JsT}).
As shown in Fig.~\ref{fig:JsTJT}, 
the value of $J_{i,i+1}^{\rm s}$ deviates from that 
for $\Delta_{\rm JT} = 0$ below $T_{\rm c}$,  
and takes two alternative values from bond to bond. 
This clearly shows 
the alternating behavior of the magnetic coupling 
in the spin-dimer state. 
At low temperatures $T \simle 150$K, 
the dimerization is almost perfect, i.e., 
$J_{i,i+1}^{\rm s}$ on the weak bonds are almost zero. 
Hence, there the system consists of almost independent spin-singlet pairs. 
Note that $J_{i,i+1}^{\rm s}$ on the strong bonds 
is surprisingly enhanced up to $\sim 1000$K. 
The almost perfect dimerization and the enhanced exchange coupling are 
the unique aspects of the present spin-orbital-lattice coupled system.  

Our estimates of the JT stabilization energy $\Delta_{\rm JT} \sim 90$K 
are considerably smaller than $T_{\rm c} = 210$K 
as well as $J = 200-300$K. 
This suggests that the critical temperature $T_{\rm c}$ is not determined 
mainly by the JT coupling, and that 
the balance of the JT coupling and the spin-orbital intersite interaction 
is important in this $t_{2g}$ electron system. 
Moreover, we note that the phase transition occurs 
below the temperature where the orbital correlation $C_{\rm o}$ 
changes its sign from negative (antiferro type) to positive (ferro type) 
as shown in Fig.~\ref{fig:corJT} (b). 
This indicates an importance of 
the interplay and the feedback effect between spin and orbital 
in the present system. 
Therefore, we conclude that the pyroxene compounds 
are typical $t_{2g}$ electron systems 
where the JT energy scale is relatively small and 
the bare interplay between spin and orbital degrees of freedom 
plays a central role in the thermodynamic properties.

\section{summary and concluding remarks}
\label{sec:summary}

In this paper, we have studied the effective spin-orbital-lattice 
coupled model which we derived 
to understand the peculiar 
phase transition to the spin-singlet state 
in $A$TiSi$_2$O$_6$ ($A$ = Na, Li). 
Using the mean-field-type analysis and 
the numerical quantum transfer matrix method, 
we have clarified that the interplay between spin and orbital degrees 
of freedom plays a central role in 
thermodynamic properties of the system.
At high temperatures, both spin and orbital correlations are 
antiferro type and compete with each other.
As temperature decreases, 
the antiferromagnetic spin correlation grows rapidly 
and yields the sign change of 
the orbital correlation from antiferro to ferro type 
so that the frustration is released.
This ferro-type orbital correlation with the Jahn-Teller coupling 
finally causes a transition to the spin-dimer and orbital-ferro ordered phase.
The feedback effect between orbital and spin degrees of freedom 
results in peculiar 
temperature dependence of the magnetic susceptibility.
We have shown that the magnetic susceptibility data 
for NaTiSi$_2$O$_6$ can be explained by our effective model 
with realistic values of parameters. 

The transition to the spin-singlet state in the present system shows 
several different aspects 
from conventional spin-Peierls systems. 
One is the temperature dependence of the magnetic susceptibility. 
In the present system, the rapid decay of the susceptibility 
due to the spin-singlet formation occurs even without a broad peak 
as a fingerprint of well-developed spin correlations. 
This is because the driving force of the transition 
is not the magnetoelastic interaction but 
the orbital-ferro correlations assisted by the Jahn-Teller distortion. 
However, this does not mean that the orbital-lattice physics is dominant 
as in $e_g$ electron systems such as CMR manganites.
In the present $t_{2g}$ electron system, the Jahn-Teller energy scale is 
considerably smaller than the orbital and spin exchange interactions, 
and the orbital-ferro correlation is induced 
by the keen competition between the orbital and spin degrees of freedom. 
These illuminate a unique feature of the present $t_{2g}$ system, 
namely, there the interplay between orbital and spin appears explicitly 
without being dominated by Jahn-Teller physics. 

Another peculiar aspect of the spin-dimer state in our model is that 
the spin-singlet pairs are formed on the longer Ti-Ti bonds 
rather than the shorter ones. 
In the present system, say, the $d_{xy}$ orbital ordering is concomitant 
with the flattening of TiO$_6$ octahedra in the $z$ direction 
which elongates Ti-Ti bonds in the $xy$ plane. 
Since the spin exchange interaction is strong 
between the nearest-neighbor sites 
in the $xy$ plane in this $d_{xy}$ ordered state, 
the spin-singlet dimers are formed on the longer Ti-Ti bonds. 
This aspect is clearly different from the conventional spin-Peierls systems 
in which the spin-singlet pairs are on shorter bonds. 
In our model, however, we take account of only the tetragonal 
Jahn-Teller mode which couples to $d_{xy}$ and $d_{yz}$ orbitals.
For detailed comparisons with the experimental data 
which show much complicated lattice structure at low temperatures, 
it is necessary to include 
more general lattice distortions in our theory. 
In particular, we note that the magnetoelastic coupling can be substantial 
in the low-temperature orbital-ordered phase 
since the orbital-ferro ordering largely enhances 
the effective spin exchange coupling on strong bonds 
up to $\sim 1000$K as shown in Sec.~\ref{sec:comparison}. 
The magnetoelastic coupling will cause 
an opposite effect on the Ti-Ti bond lengths 
since it tends to shorten the spin-singlet Ti-Ti bonds. 
Further study is necessary to conclude the low-temperature lattice structure. 
Note that the inclusion of such magnetoelastic effect 
does not alter our conclusions 
on the mechanism of the phase transition 
because it becomes important only when the orbital-ferro ordering 
is well established far below the critical temperature.

In the present study, we have compared our results of 
the magnetic susceptibility with 
the experimental data for the Na compound NaTiSi$_2$O$_6$. 
There is another compound in this pyroxene family, i.e., LiTiSi$_2$O$_6$. 
The Li compound also shows the phase transition 
at $T_{\rm c} = 230$K 
showing a sudden decay of the magnetic susceptibility below $T_{\rm c}$. 
\cite{Isobe2002}
However, the susceptibility data show 
some extra anomalies probably due to impurity phases. 
We believe that our model describes essential physics 
in both the Na and Li compounds. 
Further experimental study including the sample refinement 
is desired to compare the data of Li compounds to our results. 

As shown in Sec.~\ref{sec:gs} and in Appendix B, 
the low-temperature phase of our effective model 
turns into the spin-ferro and orbital-antiferro ordered state 
for larger values of $\eta$ than the critical value 
of $\eta_{\rm c} \simeq 0.18$. 
The parameter $\eta$ is the ratio of the Hund's-rule coupling to 
the intraorbital Coulomb repulsion, and 
is considered to be $\eta \sim 0.1 < \eta_{\rm c}$ 
in the present compounds. 
Unfortunately, it is difficult to control the parameter $\eta$ experimentally, 
however, it may vary for different compounds to some extent. 
If $\eta$ becomes close to $\eta_{\rm c}$ in some compound, 
there would occur interesting phenomena 
related to the criticality of the phase transition at $\eta_{\rm c}$. 
One interesting example is the phase transition 
from the spin-dimer and orbital-ferro state to 
the spin-ferro and orbital-antiferro state 
by applying the external magnetic field. 
We have investigated this issue in our effective model, 
and indeed observed the field-induced phase transition. 
The results will be reported elsewhere. 
The experimental study of this issue is left for future study.

\acknowledgments
We would like to thank M.\ Isobe for useful discussions and 
providing us the experimental data. 
We also appreciate H.\ Seo for valuable comments and 
drawing our attention to the present problem 
in the early stage of the study. 
This work is supported by NAREGI Nanoscience Project.

\appendix
\section{Quantum Transfer Matrix method}
In this appendix, we briefly review the algorithm 
of the QTM method.\cite{Betsuyaku1984} 
The QTM calculation is applied to the effective 1D model (\ref{eq:tildeH})
in the self-consistent scheme as described in Sec.~\ref{sec:method}. 

After the Suzuki-Trotter decomposition, 
the partition function is represented in terms of the transfer matrix as
\begin{equation}
Z = {\rm Tr}~ e^{-\beta \tilde{\mathcal{H}}} = 
\lim_{M \to \infty} {\rm Tr}~ \prod_{n=1}^{L/2} \mathcal{T}_M
\label{eq:Z}
\end{equation}
where $\beta = 1/T$ (we set the Boltzmann constant $k_{\rm B} = 1$), 
$M$ is the Trotter number, and $L$ is the system size. 
The transfer matrix $\mathcal{T}_M$ is given by
\begin{equation}
\mathcal{T}_M = \left[ 
e^{-\beta \tilde{h}_{2n-1,2n}/M} e^{-\beta \tilde{h}_{2n,2n+1}/M} \right]^M,
\label{eq:TM}
\end{equation}
where $\tilde{\mathcal{H}}$ is decomposed into the summation of
the local Hamiltonian $\tilde{h}_{i,i+1}$ for bond $(i,i+1)$.
We omitted the index $n$ on the transfer matrix $\mathcal{T}_M$
since $\mathcal{T}_M$ for the present system is invariant 
under the two-site translation.

The advantage of the QTM method is that we can calculate 
thermodynamic quantities directly from the largest eigenvalue 
$\lambda_{\rm max}$ and the corresponding right $|v^r \rangle$ 
and left $\langle v^l |$ eigenvectors of $\mathcal{T}_M$.\cite{Suzuki1987}
(Note that the eigenvectors $|v^r \rangle$ and $\langle v^l |$ are 
different in general since $\mathcal{T}_M$ is an asymmetric matrix.)
The partition function of the infinite system is found to be 
$Z = \lim_{L \to \infty} \lambda_{\rm max}^{L/2}$, and consequently, 
the free energy per site can be obtained as 
\begin{equation}
f = -\frac{1}{2\beta} \ln \lambda_{\rm max}.
\end{equation}
Hence, once the value of $\lambda_{\rm max}$ is obtained, 
we can calculate any bulk quantity in the thermodynamic limit 
by taking the appropriate derivative of the free energy $f$.

Furthermore, we can calculate expectation values of site 
and bond operators.\cite{Wang1997}
For instance, an expectation value of a product of operators 
$\mathcal{O}_{2n-1,2n} \mathcal{O}_{2n,2n+1}$
defined at sites $(2n-1,2n,2n+1)$ is obtained from the formula
\begin{equation}
\langle \mathcal{O}_{2n-1,2n} \mathcal{O}_{2n,2n+1}\rangle = 
\lim_{M \to \infty} 
\frac{\langle v^l | \mathcal{T}^\mathcal{O}_M | v^r \rangle}
{\lambda_{\rm max}},
\label{eq:expectO}
\end{equation}
where
\begin{equation}
\mathcal{T}^\mathcal{O}_M =
\mathcal{O}_{2n-1,2n} \mathcal{O}_{2n,2n+1} \mathcal{T}_M.
\label{eq:TMO}
\end{equation}

In the present study, we diagonalize the transfer matrix $\mathcal{T}_M$ 
numerically using the simple power method, which is known to be stable in 
the diagonalization of asymmetric matrix.\cite{Carlon1999}
Then, using Eqs.~(\ref{eq:expectO}) and (\ref{eq:TMO}), 
we calculate the magnetization $m$ [Eq.~(\ref{eq:Sz})]
and the NN spin and orbital correlations $C_{\rm s}$ and $C_{\rm o}$ 
[Eqs.~(\ref{eq:Cspin}) and (\ref{eq:Corb})] by
\begin{eqnarray}
m &=& \frac{\langle S^z_{2n-1} \rangle + \langle S^z_{2n} \rangle}{2},
\label{eq:m in App}
\\
C_{\rm s} &=& \frac{\langle \mib{S}_{2n-1} \cdot \mib{S}_{2n} \rangle
       + \langle \mib{S}_{2n}   \cdot \mib{S}_{2n+1} \rangle}{2},
\label{eq:C_s in App}
\\
C_{\rm o} &=& \frac{\langle T_{2n-1} T_{2n} \rangle 
       + \langle T_{2n} T_{2n+1} \rangle}{2},
\label{eq:C_o in App}
\end{eqnarray}
respectively.
The NN spin correlations on odd and even bonds 
[Eqs.~(\ref{eq:Cspin_odd}) and (\ref{eq:Cspin_even})] 
are given by
\begin{eqnarray}
C_{\rm s}^{\rm odd} &=& \langle \mib{S}_{2n-1} \cdot \mib{S}_{2n} \rangle,
\label{eq:C_s_odd in App}
\\
C_{\rm s}^{\rm even} &=& \langle \mib{S}_{2n} \cdot \mib{S}_{2n+1} \rangle,
\label{eq:C_s_even in App}
\end{eqnarray}
respectively.
In the spin-dimer phase, 
$C_{\rm s}^{\rm odd}$ and $C_{\rm s}^{\rm even}$ may take different values.

\section{transition to spin-F and orbital-AF phase}
\begin{figure}
\includegraphics[width=80mm]{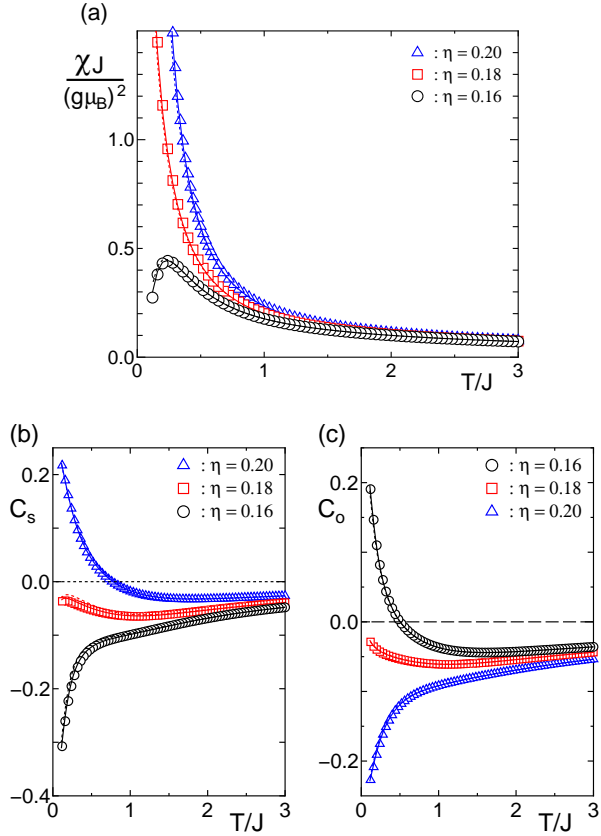}
\caption{
Temperature dependences of (a) the magnetic susceptibility $\chi$, 
(b) the nearest-neighbor spin correlation $C_{\rm s}$, 
and (c) the nearest-neighbor orbital correlation $C_{\rm o}$ 
for $\eta = 0.16, 0.18$ and $0.20$. 
Symbols represent the results for $M = 4$ while 
solid and dotted curves are those for $M = 3$ and $M = 2$, respectively.
} 
\label{fig:largeeta}
\end{figure}
In this appendix, we discuss 
thermodynamic properties of the model (\ref{eq:H}) 
for larger values of $\eta$ than those studied 
in Sec.~\ref{sec:numerical results}.
There, the sF-oAF ground state is expected to be stable
from the mean-field-type analysis in Sec.~\ref{sec:gs}.

We first discuss our numerical results for the spin-orbital model 
$\mathcal{H}_{\rm so}$, i.e., the effective model (\ref{eq:tildeH}) 
without the JT coupling $\bar{\gamma} = 0$, 
corresponding to the results in Sec.~\ref{sec:interplay}.
Figure~\ref{fig:largeeta} (a) shows the results of 
the magnetic susceptibility $\chi$.
The result for $\eta = 0.16$ exhibits a sharp drop 
as $T \to 0$ similarly to the results in Fig.~\ref{fig:chi-eta}, 
indicating the spin-singlet ground state.
On the contrary, 
for $\eta = 0.18$ and $0.20$, 
$\chi$ exhibits a divergent behavior as $T \to 0$, 
suggesting that the ground state is magnetic.
These suggest that there is a ground-state phase transition 
between nonmagnetic and magnetic phases at $\eta = \eta_{\rm c} \sim 0.18$.

To clarify the nature of the transition in more detail, 
we show the results of 
NN spin and orbital correlations $C_{\rm s}$ and $C_{\rm o}$ 
in Figs.~\ref{fig:largeeta} (b) and (c), respectively.
The results for $\eta = 0.16$ show similar features to 
those for smaller $\eta$ shown in Sec.~\ref{sec:numerical results}, 
and indicate that the system exhibits the sD-oF ground state.
For $\eta = 0.20$, on the other hand, 
the antiferro-type orbital correlation $C_{\rm o}$ develops rapidly, 
yielding the sign change of the spin correlation $C_{\rm s}$.
After the sign change, the two correlations grow cooperatively, 
and approach the values of  
$C_{\rm s} = 1/4$ and $C_{\rm o} = -1/4$ as $T \to 0$.
These are the values expected in the sF-oAF ground state 
shown in Fig.~\ref{fig:SOpattern} (b).
Hence, the system with $\eta = 0.20$ exhibits the sF-oAF ground state.
For $\eta = 0.18$, $C_{\rm s}$ and $C_{\rm o}$ are both antiferro type and 
compete with each other down to the lowest temperature studied here. 
This suggests that $\eta = 0.18$ is close to 
the phase boundary between the sD-oF and sF-oAF phases. 
Therefore, we conclude that 
the ground state of the spin-orbital model $\mathcal{H}_{\rm so}$ 
undergoes a phase transition between 
the sD-oF and sF-oAF phases at $\eta = \eta_{\rm c} \sim 0.18$.
The critical value is 
in good agreement with the mean-field prediction in Sec.~\ref{sec:gs}.

In the case of a finite JT coupling, we have found that 
there occurs a finite-temperature phase transition 
to the low-temperature sF-oAF phase for $\eta > \eta_{\rm c}$ or 
to the sD-oF phase for $\eta < \eta_{\rm c}$. 
The details will be reported elsewhere.



\begin{thebibliography}{99}
\bibitem{effectofJT}
We note that the transfer integrals 
may be modified by the JT distortion. 
We neglect the small corrections in the present study.
\bibitem{noneedg}
Note that the results of the correlations are independent of the value of $g$.
\end{thebibliography}
\end{document}